# Universal Numeric Segment Display for Indian Scheduled Languages: an Architectural View

Partha Pratim Ray

*Surendra Institute of Engineering and Management*
*Siliguri, Darjeeling-734009, West Bengal, India*

*Abstract—* **India is country of several hundred different languages. Though twenty two languages have only been devised as scheduled to the Eighth Schedule of Indian Constitution in 2007. But as there is yet no proposed compact display architecture to display all the scheduled language numerals at a time, this paper proposes a uniform display architecture to display all twenty two different language digits with higher accuracy and simplicity by using a 17-segment display, which is an improvement over the 16-segment display.**

## I. INTRODUCTION

India is country of multiple languages. Most Indians speak a language belonging either to the Indo-European (74%), the Dravidian (24%), the Austro-Asiatic (Munda) (1.2%), or the Tibeto-Burman (0.6%) families [1]. The Eighth Schedule to the Indian Constitution contains a list of 22 scheduled languages. About 95% native Indians speak in these languages. Indian Government has also decided to include all the scheduled languages for official use—including message board in airports, rail stations, bus stands and other non official areas (e.g. banking, hospitals, sports, medical, shopping malls etc.) to communicate with people. Most of these cases numerals play vital role. Hence the need for designing a display architecture which will help Indians to understand the numerals in message boards, aroused.

Segment display is now a widely used display method in electronic devices. Because numerals have lots of curve edges which is better supported by matrix display. But due to the high cost and complexity of matrix displays, segment display is generally used to display numerals. In this paper I have proposed a 17-segment display architecture which is a slight modification of conventional 16-segment display. My proposed architecture can be able to display numerals of the following scheduled Indian languages: - Assamese, Bengali, Bodo, Dogri, Gujarati, Hindi, Kannada, Konkani, Kashmiri, Maithili, Malayalam, Manipuri, Marathi, Nepali, Oriya, Punjabi, Sanskrit, Santali, Sindhi, Tamil, Telugu, Urdu along with English.

This paper is organized as follows. Section II states related work. Section III describes the proposed segment display architecture. Section IV presents various patterns of numerals. Section V concludes this paper.

## II. RELATED WORK

So far 7-segment display is used for English digits, those are Latin digits and 10-segment display is proposed for Bengali digits [2, 3]. [4] proposed an 8-segment display is English and Bengali digits. Literature [5] proposed a 12-segment display for Arabic digits. [6] described an 18-segmented display for twelve international languages. Another report, [7] presented a 31-segmented display for Bengali characters. [8] proposed a 26-segment display for Bengali characters. But these segment architectures are different to each other and we cannot use the same architecture for uniform display of multiple languages, that's why I have proposed 17-segment display as a uniform architecture to display multiple numerals at a time. My proposed 17-segment display can be used to display numerals of twenty two Indian languages along with English.

## III. PROPOSED SEGMENT DISPLAY ARCHITECTURE

In the current world 7-segment display is the all in all in Latin type numeral display but cannot be used as a universal numeric display. My proposed 17-segment display is an improvement over conventional 16-segment.

I have added one extra segment in the 16-segment display— *p segment*. The segment p is placed in the lower layer part of the display. Finally the segment display takes the form of a 17-segment display. The 17-segment display architecture is shown below, Fig. 1 shows the traditional 16-segment display and the Fig. 2 shows the proposed 17-segment display. My 17-segment display requires one more segment in comparison to the common 16-segment display. As 16-segment display is very common in practice and quite cheap and needs fewer gates to implement that is why I have proposed to use 17-segment display with a bit modification. Moreover by using 17-segment displays we may eliminate the need of a completely new specially fabricated segment display. By adding one segment 16-segment display can be easily converted to 17-segment display.

The pattern of different language numeral are shown in the below tabulated figures (Fig. 3 – Fig. 19). In the following figures "D Val" stands for *Digit Value*, "Act Sym" stands for *Actual Symbol*, "Seg Pat" stands for *Segment Pattern*, and "Com Vec" stands for *Combination Vector*.

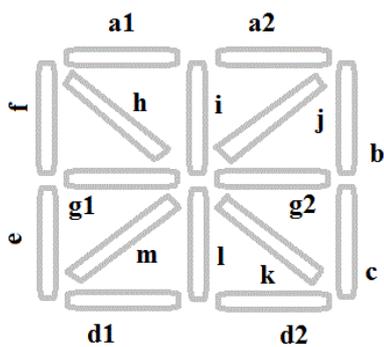

Fig. 1  16-segment display.

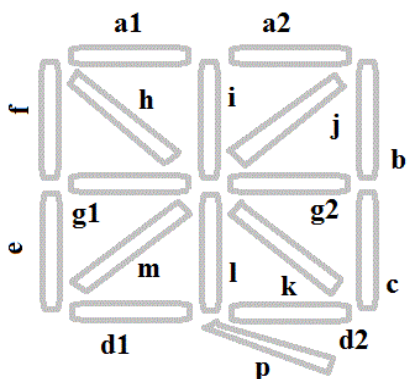

Fig. 2  Proposed 17-segment display.

IV. PATTERNS OF NUMERALS

Following figures presents the patterns of different language numerals.

| D Val | Act Sym | Seg Pat | Com Vec |
|---|---|---|---|
| 0 | ০ | | { a1, a2, b, c, d2, d1, e, f } |
| 1 | ১ | | { a1, a2, b, c, d2, d1, e, g1, l } |
| 2 | ২ | | { a1, a2, b, d2, d1, e, g1, g2 } |
| 3 | ৩ | | { a2, b, c, d2, d1, e, f, i, g2 } |
| 4 | ৪ | | { a1, a2, b, c, d2, d1, e, f, g1, g2 } |
| 5 | ৫ | | { a1, c, d2, d1, e, f, i, g2 } |
| 6 | ৬ | | { i, g2, c, d2, d1, e, f } |
| 7 | ৭ | | { a1, a2, b, c, f, g1, g2 } |
| 8 | ৮ | | { d1, e, f, g1, l, j } |
| 9 | ৯ | | { a1, a2, b, c, d2, e, g1, l } |

Fig. 3  Representation of Bengali and Assamese numerals

| D Val | Act Sym | Seg Pat | Com Vec |
|---|---|---|---|
| 0 | . | | { a1, f, i, g1 } |
| 1 | ੧ | | { a1, e, f, i, l } |
| 2 | ੨ | | { a1, d1, g1, i, l } |
| 3 | ੩ | | { a1, d1, p, g1, i, l } |
| 4 | ੪ | | { d1, h, j, l, m } |
| 5 | ੫ | | { f, i, g1, l } |
| 6 | ੬ | | { a1, h, i, l, g1 } |
| 7 | ੭ | | { a2, b, c, i, g2 } |
| 8 | ੮ | | { a1, f, g1, l, d1 } |
| 9 | ੯ | | { a1, a2, d1, e, f, g1, l } |

Fig. 4  Representation of Dogri numerals

| D Val | Act Sym | Seg Pat | Com Vec |
|---|---|---|---|
| 0 | ૦ | | { a1, a2, b, c, d1, d2, e, f } |
| 1 | ૧ | | { a1, p, f, g1, i, l } |
| 2 | ૨ | | { a2, b, g2, k } |
| 3 | ૩ | | { a1, a2, b, c, d2, d1, g2 } |
| 4 | ૪ | | { d1, h, j, l, m } |
| 5 | ૫ | | { a1, b, c, i, g2 } |
| 6 | ૬ | | { a1, a2, d1, e, f, g1, g2 p } |
| 7 | ૭ | | { a2, b, c, d2, d1, e, f, i } |
| 8 | ૮ | | { d2, d1, j, m } |
| 9 | ૯ | | { d2, d1, e, f, j, m } |

Fig. 5  Representation of Gujarati numerals

| D Val | Act Sym | Seg Pat | Com Vec |
|---|---|---|---|
| 0 | ० | | {a1, a2, b, c, d2, d1, e, f} |
| 1 | १ | | {a1, p, d1, f, g1, i, m} |
| 2 | २ | | {a1, p, d1, i, l} |
| 3 | ३ | | {a1, p, d1, g1, i, l} |
| 4 | ४ | | {d1, h, j, l, m} |
| 5 | ५ | | {p, d1, e, f, l} |
| 6 | ६ | | {a1, p, d1, e, f, g1} |
| 7 | ७ | | {a2, b, c, d2, d1, e, f, i, g2} |
| 8 | ८ | | {d2, d1, j, m} |
| 9 | ९ | | {a2, b, d2, i, g2, k} |

Fig. 6  Representation of Bodo, Hindi, Konkani, Marathi, Nepali and Sanskrit numerals

| D Val | Act Sym | Seg Pat | Com Vec |
|---|---|---|---|
| 0 | ೦ | | {a1, a2, b, c, d2, d1, e, f} |
| 1 | ೧ | | {a1, a2, b, c, d1, e, f} |
| 2 | ೨ | | {a2, b, c, d2, d1, i, j} |
| 3 | ೩ | | {a1, p, d1, i, m} |
| 4 | ೪ | | {a1, d1, h, j, l, m} |
| 5 | ೫ | | {a1, b, d1, i, j, g2, k, l, m} |
| 6 | ೬ | | {d2, d1, e, f, g1} |
| 7 | ೭ | | {a2, b, d2, d1, g2, m} |
| 8 | ೮ | | {a2, d1, e, f, i, l} |
| 9 | ೯ | | {a1, a2, d1, e, f, g1} |

Fig. 7  Representation of Kannada numerals

| D Val | Act Sym | Seg Pat | Com Vec |
|---|---|---|---|
| 0 | ٠ | | {a1, f, i, g1} |
| 1 | ١ | | {e, f} |
| 2 | ٢ | | {e, f, g1, i} |
| 3 | ٣ | | {e, f, g1, g2, i, b} |
| 4 | ٤ | | {a1, d1, h, m} |
| 5 | ٥ | | {d2, d1, e, f, h, k} |
| 6 | ٦ | | {b, c, i, g2} |
| 7 | ٧ | | {b, c, h, k} |
| 8 | ٨ | | {b, c, j, m} |
| 9 | ٩ | | {a2, b, c, i, g2} |

Fig. 8  Representation of Kashmiri numerals

| D Val | Act Sym | Seg Pat | Com Vec |
|---|---|---|---|
| 0 | ० | | {a1, a2, b, c, d2, d1, e, f} |
| 1 | १ | | {a1, a2, b, c, d2, d1, e, g1, l} |
| 2 | २ | | {a1, a2, b, d1, d2, e, g1 g2} |
| 3 | ३ | | {a2, b, c, d2, g2, p} |
| 4 | ४ | | {d1, h, j, l, m} |
| 5 | ५ | | {a1, a2, c, d2, d1, e, i, g2} |
| 6 | ६ | | {c, d2, d1, e, f, i, g2} |
| 7 | ७ | | {a2, b, c, i} |
| 8 | ८ | | {i, g2, d1, l, m} |
| 9 | ९ | | {b, c, d2, e, g1, l} |

Fig. 9  Representation of Maithili numerals

| D Val | Act Sym | Seg Pat | Com Vec |
|---|---|---|---|
| 0 | ൦ | | {a1, a2, d1, e, f, g1, l } |
| 1 | ൧ | | {a2, b, d1, i, g2, m } |
| 2 | ൨ | | {d2, e, g1, l } |
| 3 | ൩ | | {c, e, g1, g2, l } |
| 4 | ൪ | | {a2, b, c, d2, e, g1, g2, l } |
| 5 | ൫ | | {a1, a2, b, d1, e, f, h, i, g1, l } |
| 6 | ൬ | | {a1, a2, b, c, e, f, i, l } |
| 7 | ൭ | | {a1, a2, b, d1, e, f, l, m } |
| 8 | ൮ | | {b, c, d2, e, g1, l, p } |
| 9 | ൯ | | {a2, b, c, d2, e, f, g1, g2, l } |

Fig. 10  Representation of Malayalam numerals

| D Val | Act Sym | Seg Pat | Com Vec |
|---|---|---|---|
| 0 | ୦ | | {a1, a2, b, c, d2, d1, e, f } |
| 1 | ୧ | | {a1, d1, e, f, g1, i } |
| 2 | ୨ | | {a2, b, c, d2, i, g2 } |
| 3 | ୩ | | {a1, a2, b, c, f, i, l, g1 } |
| 4 | ୪ | | {h, j, k, m, d2, d1 } |
| 5 | ୫ | | {a2, d1, i, j, l, m, g1, g2 } |
| 6 | ୬ | | {a2, b, c, d2, d1, h, i } |
| 7 | ୭ | | {a1, a2, b, c, d2, i, j } |
| 8 | ୮ | | {a1, e, f } |
| 9 | ୯ | | { d2, d1, e, f, h, i } |

Fig. 12  Representation of Oriya numerals

| D Val | Act Sym | Seg Pat | Com Vec |
|---|---|---|---|
| 0 | ꯰ | | {a1, a2, b, c, d2, d1, e, f } |
| 1 | ꯱ | | {a1, a2, d1, f, g1, l } |
| 2 | ꯲ | | {a1, a2, d1, f, g1, l, m } |
| 3 | ꯳ | | {a1, a2, c, d1, f, g1, g2, k, l, m } |
| 4 | ꯴ | | {a1, d2, d1, e, g1, i } |
| 5 | ꯵ | | { a1, a2, f, g1, l } |
| 6 | ꯶ | | {a1, a2, f, d2 g1, l } |
| 7 | ꯷ | | {a1, b, d1, i, j, g2, k , l, m } |
| 8 | ꯸ | | {a1, a2, b, d1, f, g1, l } |
| 9 | ꯹ | | {a1, a2, b, d1, f, g1, k, l, m} |

Fig. 11  Representation of Manipuri numerals

| D Val | Act Sym | Seg Pat | Com Vec |
|---|---|---|---|
| 0 | ੦ | | {a1, a2, b, c, d2, d1, e, f } |
| 1 | ੧ | | {a1, p, f, i, g1, l } |
| 2 | ੨ | | {a1, p, d1, i, l } |
| 3 | ੩ | | {a1, p, d1, i, g1, l } |
| 4 | ੪ | | {h, j, d1, l, m } |
| 5 | ੫ | | {f, i, l, g1 } |
| 6 | ੬ | | {a1, a2, d1, e, f, g1 } |
| 7 | ੭ | | {a2, b, c, d2, d1, i } |
| 8 | ੮ | | {a1, d2, d1, e, f } |
| 9 | ੯ | | {a1, d2, d1, e, f, h } |

Fig. 13  Representation of Punjabi numerals

| D Val | Act Sym | Seg Pat | Com Vec |
|---|---|---|---|
| 0 | ০ | | {a1, a2, b, c, d2, d1, e, f} |
| 1 | ১ | | {d2, d1, e, f, h, i} |
| 2 | ২ | | {a1, d2, d1, e, h, i, g1} |
| 3 | ৩ | | {d2, d1, f, g1, h, i, m} |
| 4 | ৪ | | {a1, d1, e, f, g1, l} |
| 5 | ৫ | | {d2, d1, e, f, j, k} |
| 6 | ৬ | | {a1, d2, d1, e, f, h, i} |
| 7 | ৭ | | {a1, d2, d1, e, i, g1} |
| 8 | ৮ | | {b, d2, d1, f, h, j, g2, m} |
| 9 | ৯ | | {a1, c, d2, d1, e, f, g2} |

Fig. 14  Representation of Santali numerals

| D Val | Act Sym | Seg Pat | Com Vec |
|---|---|---|---|
| 0 | O | | {a1, a2, b, c, d2, d1, e, f} |
| 1 | ௧ | | {a1, a2, c, e, f, g1, g2, i, m} |
| 2 | ௨ | | {a1, d2, d1, e, f, i, g1} |
| 3 | ௩ | | {a1, a2, c, d2, p, e, f, i, g2} |
| 4 | ௪ | | {a1, a2, b, d1, e, f, i, l, g1, g2} |
| 5 | ௫ | | {a1, a2, c, e, f, d2, d1, i, l, g2} |
| 6 | ௬ | | {a1, a2, c, e, f, i, l, m, g1, g2} |
| 7 | ௭ | | {a1, a2, b, c, d1, e, f, l, g1} |
| 8 | ௮ | | {a1, b, c, d1, e, h, i, l, g1, g2} |
| 9 | ௯ | | {a1, a2, c, d1, e, f, i, k, l, g1, g2} |
| 10 | ௰ | | {b, c, d2, d1, e, f, i, l} |
| 100 | ௱ | | {a1, a2, b, c, e, f, i, l} |
| 1000 | ௲ | | {a1, a2, c, d2, p, e, f, i, g1, g2, m} |

Fig. 16  Representation of Tamil numerals

| D Val | Act Sym | Seg Pat | Com Vec |
|---|---|---|---|
| 0 | ٠ | | {a1, a2, b, c, d2, d1, e, f} |
| 1 | ١ | | {a1, p, f, i, l, g1} |
| 2 | ٢ | | {a2, b, g2, k} |
| 3 | ٣ | | {a2, b, c, d2, g2} |
| 4 | ٤ | | {d1, h, j, l, m} |
| 5 | ٥ | | {p, f, i, g1, l} |
| 6 | ٦ | | {a1, d1, g1, e, f} |
| 7 | ٧ | | {a2, b, c, g2, i, d2, d1, e, f} |
| 8 | ٨ | | {d2, d1, j, m} |
| 9 | ٩ | | {d2, d1, e, f, j, m} |

Fig. 15  Representation of Sindhi numerals

| D Val | Act Sym | Seg Pat | Com Vec |
|---|---|---|---|
| 0 | ౦ | | {a1, a2, b, c, d2, d1, e, f} |
| 1 | ౧ | | {a1, a2, b, c, d1, e, f} |
| 2 | ౨ | | {a2, b, c, d2, d1, i, g2} |
| 3 | ౩ | | {a2, b, c, d2, g2} |
| 4 | ౪ | | {b, d2, d1, f, g1, g2, k, m} |
| 5 | ౫ | | {a2, c, h, i, j, m, g2} |
| 6 | ౬ | | {a1, d2, d1, e, f, g1} |
| 7 | ౭ | | {a1, a2, b, d2, d1, e, g1, g2} |
| 8 | ౮ | | {a2, d1, e, f, i, l} |
| 9 | ౯ | | {a1, a2, d1, e, f, g1} |

Fig. 17  Representation of Telugu numerals

| D Val | Act Sym | Seg Pat | Com Vec |
|---|---|---|---|
| 0 | ٠ | | { a1, i, g1, f } |
| 1 | ١ | | { e, f } |
| 2 | ٢ | | { e, f, g1, i } |
| 3 | ٣ | | { b, e, f, g1, g2, i } |
| 4 | ۴ | | { a1, a2, b, e, f, i } |
| 5 | ۵ | | { d2, d1, e, f, h, k, m } |
| 6 | ۶ | | { b, c, i, g2 } |
| 7 | ۷ | | { d2, d1, j, m } |
| 8 | ۸ | | { b, c, j, m } |
| 9 | ۹ | | { a1, f, i, l, g1 } |

Fig. 18  Representation of Urdu numerals

| D Val | Act Sym | Seg Pat | Com Vec |
|---|---|---|---|
| 0 | 0 | | { a2, b, c, d2, i, l } |
| 1 | 1 | | { b, c } |
| 2 | 2 | | { a2, b, d2, g2, l } |
| 3 | 3 | | { a2, b, c, d2, g2 } |
| 4 | 4 | | { b, c, i, g2 } |
| 5 | 5 | | { a2, c, d2, i, g2 } |
| 6 | 6 | | { a2, c, d2, i, g2, l } |
| 7 | 7 | | { a2, b, c } |
| 8 | 8 | | { a2, b, c, d2, i, g2, l } |
| 9 | 9 | | { a2, b, c, i, g2 } |

Fig. 19  Representation of English numerals

## V. CONCLUSIONS

In this paper I have proposed 17-segment display for representing numerals of twenty two different language numeric symbols of India. As this display architecture supports multiple language numerals together, it can be considered as the simplest universal display. In future I will try to improve its usability by displaying numerals of other languages.